\documentclass{aa}

\usepackage{graphicx}
\usepackage{rotating} 
\usepackage{psfig}
\usepackage{latexsym}
\usepackage{amssymb}
\usepackage{natbib}
\usepackage[innercaption]{sidecap}

\def\cm3{${\rm cm^{-3}}$}            
\def\mum{${\rm \, \mu m}$}                            
\def\kms{\hbox{\,km\,s$^{-1}$}}  

\newcommand{\Teff}{$T_{\rm eff}$}                   
\newcommand{\Tel}{$T_{\rm e}$}                      
\newcommand{\nel}{$n_{\rm e}$}                      
\newcommand{\rgal}{$R_{\rm Gal}$}

\newcommand{\HI}{\ion{H}{i}}
\newcommand{\HII}{\ion{H}{ii}}
\newcommand{\HeI}{\ion{He}{i}}
\newcommand{\HeII}{\ion{He}{ii}}
\newcommand{\FeIII}{[\ion{Fe}{iii}]}
\newcommand{\ArII}{[\ion{Ar}{ii}]}
\newcommand{\ArIII}{[\ion{Ar}{iii}]}
\newcommand{\NeII}{[\ion{Ne}{ii}]}
\newcommand{\NeIII}{[\ion{Ne}{iii}]}
\newcommand{\SIII}{[\ion{S}{iii}]}
\newcommand{\SIV}{[\ion{S}{iv}]}
\newcommand{\NII}{[\ion{N}{ii}]}
\newcommand{\NIII}{[\ion{N}{iii}]}

\begin{document}

   \title{A VLT spectroscopic study of the ultracompact \HII\ region 
  G29.96$-$0.02}
   
   \author{
    N.\,L. Mart\'{\i}n-Hern\'{a}ndez\inst{1}\thanks{\emph{Present address:}
                 Geneva Observatory, 1290 Sauverny, Switzerland}
    \and A. Bik\inst{2}
    \and L. Kaper\inst{2}
    \and A.\,G.\,G.\,M. Tielens\inst{1,3}
    \and M.\,M. Hanson\inst{4}
   }

   \offprints{N.L.\,Mart\'{\i}n-Hern\'{a}ndez, 
     email: leticia.martin@obs.unige.ch}

   \institute{
Kapteyn Institute, P.O. Box 800, 9700 AV Groningen, The Netherlands
   \and
Astronomical Institute Anton Pannekoek, University of Amsterdam, 
Kruislaan 403, 1098 SJ Amsterdam, The Netherlands
   \and
SRON, National Institute for Space Research, P.O. Box 800,
9700 AV Groningen, The Netherlands 
   \and 
Department of Physics, University of Cincinnati, Cincinnati, OH 45221, USA
   }

   \date{}

\abstract{
A high quality, medium-resolution K-band spectrum has been
obtained of the ultracompact \HII\ region G29.96$-$0.02 with the {\it Very
Large Telescope} (VLT). The slit was positioned along the symmetry axis of
the cometary shaped nebula. Besides the spectrum of the embedded
ionizing O star, the long-slit observation revealed the rich
emission-line spectrum produced by the ionized nebula with sub-arcsec
spatial resolution. The nebular spectrum includes Br$\gamma$, several
helium emission lines and a molecular hydrogen line. A detailed
analysis is presented of the variation in strength, velocity and width
of the nebular emission lines along the slit. The results are
consistent with previous observations, but the much better spatial
resolution allows a critical evaluation of models explaining the
cometary shape of the nebula. Our observations support neither
the wind bow shock model nor the
champagne flow model.\\ The measured line ratios of the nebular
hydrogen and helium lines are compared to predictions from case~B
recombination-line theory. The results indicate an electron
temperature between 6400 and 7500 K, 
in good agreement with other determinations
and the Galactocentric distance of 4.6~kpc. The He$^{+}$/H$^{+}$ ratio
is practically constant over the slit; we argue that He is singly
ionized throughout the nebula. We review the various observational
constraints on the effective temperature of the
ionizing star and show that these are in agreement with its K-band
spectral type of O5--O6\,V.
\keywords{Infrared: ISM -- ISM: lines and bands -- Stars: early type -- 
\HII\ regions -- ISM: individual: G29.96--0.02 -- 
ISM: kinematics and dynamics}
}

\maketitle

\section{Introduction}

\HII\ regions trace the formation sites of massive stars. The
ionizing radiation of the embedded, young OB stars is absorbed by gas
and dust in their near surroundings and re-emitted at infrared and
radio wavelengths. Their progenitors, the ultracompact \HII\ 
(UCH{\sc ii}) regions,
represent the earliest recognizable phase of massive-star
formation, and are the most luminous sources at 100~$\mu$m, observable
throughout the Galaxy \citep{churchwell90,garay99}.  The
small size and large number of UCH{\sc ii} regions 
indicate that the lifetime of
this phase is significantly longer than the $\sim 10^{4}$~year
predicted by models describing the expansion of young \HII\ regions. 

A possible solution for this ``confinement problem'' is suggested by
the observed morphologies of UCH{\sc ii} regions. The frequently seen cometary
structures are suggestive of bow shocks. \cite{wood89}
proposed that these bow shocks might result from the supersonic motion
($v_{\rm sound} \simeq 1$~\kms) of the embedded O stars relative to
the ambient molecular cloud. The stand-off distance of the shock, and
thus the observed ``size'' of the UCH{\sc ii} regions, 
is determined by the density
of the ambient medium, the velocity of the O star, and the momentum of
its stellar wind. These parameters are not expected to vary rapidly
with time, thus delaying the apparent expansion rate of the UCH{\sc ii}
region \citep[e.g.][]{vanburen90,maclow91,vanburen92}.

G29.96$-$0.02, also denominated IRAS 18434$-$0242 (hereafter G29.96), 
is a well-studied,
cometary-shaped  UCH{\sc ii} region at a Galactocentric distance of
4.6~kpc \cite[e.g.][]{pratap99}. 
Arguments based on extinction and the spectral
type of the ionizing star favour the near heliocentric distance of
6~kpc instead of the far distance of 9 kpc \citep{pratap99}.  The
cometary structure of G29.96 was first observed by \cite{wood89} at
radio wavelengths.  Follow-up radio observations were performed by
\cite{wood91}, \cite{afflerbach94}, \cite{cesaroni94}, \cite{fey95} 
and \cite{kim01}. The cometary shape is characterized by a sharp,
bright, arc-like leading edge and a low-surface brightness tail of
emission which trails off away from the edge. \cite{wood89} found the
bow shock hypothesis consistent with their radio recombination line
observations of G29.96. Alternatively, a champagne-flow model, which
describes the spherical expansion of an \HII\ region into a molecular
medium with a steep density gradient \citep{yorke86}, has been
proposed to explain the morphology and kinematical structure of G29.96
\citep{lumsden96,lumsden99}. 

G29.96 is embedded in a molecular cloud that produces $\sim 15$ magnitudes
of visual extinction \citep[see][]{martin:paperii}. 
This molecular cloud has
been extensively studied at low spatial resolution
\citep{churchwell90b,churchwell92}. Strong
emission from different dense gas tracers indicates the presence of
hot, dense gas. Just in front of the edge of the \HII\ region, at
approximately 2\arcsec\ west of the radio continuum peak, a hot
molecular core was discovered in interferometric observations
\citep{cesaroni94,cesaroni98,maxia01}. Recently, this hot core has
been found at mid-infrared wavelengths too \citep{deBuizer02}. The
observed offset between the hot core and the \HII\ region and the
detection of masers towards the core
\citep{pratap94,hofner96,walsh98} suggest that the hot core is not
heated by the ionizing star of G29.96, but by another young, deeply
embedded massive star.

G29.96 seems to be associated with a small cluster of stars
\citep{fey95,lumsden96,watson97a,pratap99}. The main ionizing star is 
located in the focus of the bright arc.
\cite{watson97b} obtained a K-band spectrum of the embedded ionizing
star, the first ionizing star of an UCH{\sc ii} region 
ever detected. Based upon
the presence and strength of photospheric lines of \ion{C}{iv} and
\ion{N}{iii}, they deduced a main sequence spectral type between O5 
and O8. 

G29.96 provides an excellent laboratory to test models describing the
ionization structure of an UCH{\sc ii} region. The ionization
structure can be traced by the nebular hydrogen and helium
recombination lines. The comparison between observations and model
provides an important constraint on the stellar parameters of the
ionizing star(s), yielding important information on the properties of
newly formed massive stars
\citep[e.g.][]{vanzi96,lumsden96b,lumsden01,hanson02}. Due to the
large visual extinction towards \HII\ regions like G29.96, such a
study has only recently become possible thanks to the technological
development of near-IR spectrometers.  

In this paper we present new,
high quality, long-slit K-band spectra of G29.96
(2.07--2.19~\mum). Besides Br$\gamma$, a large number of \HeI\ lines
and a H$_2$ line are covered. The study of the relative strengths of
the \HI\ and \HeI\ lines as a function of position along the slit
allows different lines of investigation. First, it is possible to
study the impact of large line opacities on the \HeI\ lines. Second,
the conditions present in the ionized gas can be determined. 
Third, the results constrain the
effective temperature of the ionizing star, which can be compared with
the spectral type implied by its K-band spectrum. Finally, the
kinematical structure of the ionized gas can be studied at
higher spatial resolution than in previous studies
\citep{wood91,afflerbach94,lumsden96,lumsden99} and compared with 
the molecular gas motions. This
comparison can be used to test the different models (bow shock,
champagne flow, etc.) that have been proposed to explain the cometary
morphology of G29.96.

   \begin{SCfigure*}[1.0][!ht] 
     \centering
     \includegraphics[width=10cm]{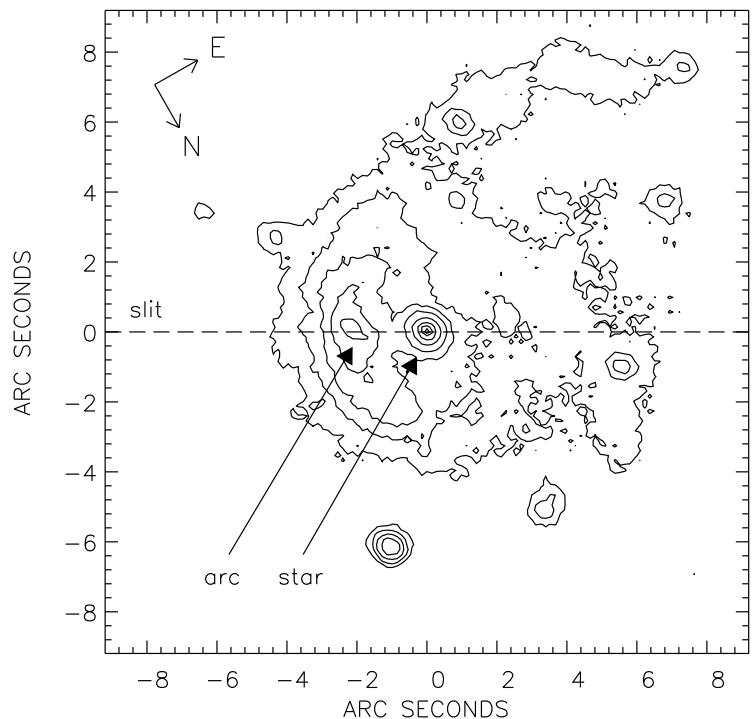} 
     \caption{Acquisition image of the UCH{\sc ii} region
       G29.96--0.02 obtained through a narrow-band H$_2$ 1--0 S(1)
       filter (central wavelength 2.13 \mum). Coordinates are relative
       to the position of the ionizing star: 
       $\alpha(2000)=18^{\rm h}46^{\rm m}4\fs075, 
       \delta(2000)=-2\degr 39\arcmin 21\farcs42$.
       Contours are 1, 2.5, 5, 10, 25, 50, 75 and 90\% the peak flux.
       The cometary structure of the \HII\ region is
       clearly revealed.
       The position of the slit is marked by a dashed line. 
       The ionizing star and the bright arc are
       indicated. At a distance of 6~kpc, 1~arcsec corresponds to
       0.029 pc.}
     \label{fig:vlt:G29}
   \end{SCfigure*}

This paper is structured as follows. Section~\ref{sect:vlt:obs}
describes the observations and the data
reduction. Section~\ref{sect:vlt:results} presents the resulting spectra
and the variations and kinematics of the lines across
G29.96. Section~\ref{sect:vlt:theory} describes the \HI\ and \HeI\
recombination theory and shows the different
applications of these lines. 
Section~\ref{sect:vlt:sptype} reviews the various observational
constraints on the effective temperature of the ionizing star.
Section~\ref{sect:vlt:dynamics} discusses the dynamics of G29.96.
Finally, Sect.~\ref{sect:vlt:summary} summarizes the results of the paper.

\section{Observations}
\label{sect:vlt:obs}

Long-slit (120\arcsec) K-band spectra were obtained of the UCH{\sc ii}
region G29.96 with ISAAC mounted on {\it Antu} (UT1) of ESO's 
{\it Very
Large Telescope} (VLT), Paranal, Chile. The observations were carried
out on 19 March 2000.  A slit-width of 0\farcs3 was used, resulting
in a spectral resolving power R$\sim 8000$ ($=$c/$\Delta$v). The slit
was positioned along the symmetry axis of the object (116.2 degrees
with respect to the north-south axis). Figure~\ref{fig:vlt:G29} shows
the location of the slit in the (acquisition) image of G29.96 obtained
through a narrow-band H$_2$ filter (central wavelength 2.13~\mum).

The observing conditions at Paranal Observatory were excellent
(humidity less than 10\%\ and seeing of 0\farcs6). 
In order to correct for the sky background, the object
was ``nodded'' between two positions on the slit (A and B) such that
the background emission registered at position B (when the source is
at position A) is subtracted from the source plus sky background
observation at position B in the next frame, and vice versa. This
strategy has the advantage that observing time is very efficiently
used, but it reduces the effective length of the slit by about
50\%; for G29.96 this is not a problem.

The electrical ghosts and bias were removed from the frames before
flatfielding. Subsequently, the sky background emission was subtracted
following the procedure outlined above. Wavelength calibration was
performed using the telluric OH lines. The accuracy of the wavelength
calibration is 3~\kms. To correct for absorption lines due to the
Earth's atmosphere, the target spectra were divided by the spectrum of
a telluric standard star. For this purpose,  an A-type star was observed
under identical sky conditions, which provides a continuous
spectrum only containing telluric absorption lines. The hydrogen
Br$\gamma$ line in the A-star spectrum was removed by a model fit. The
telluric standard was also used to correct for the throughput of
telescope and instrument. For a more detailed description of the data
reduction procedures we refer to Bik et al. (in preparation).

Spectra were extracted along the slit (1 pixel = 0\farcs147) in the
spatial direction. The peak positions, equivalent widths and fluxes of
the lines were measured by fitting a Voigt profile.  A line is defined
as being detected if its peak intensity exceeds the rms noise level of
the local continuum by at least a factor three. The central wavelength
of the line was measured with respect to the local standard of rest,
i.e. corrected for the projected velocity of the Earth with respect to
the Sun at the time of observation and for the solar motion in the
direction of G29.96 ($v_{\rm LSR}=44$~\kms).

   \begin{figure*}[!ht]
   \centering
   \includegraphics[height=13cm,angle=90]{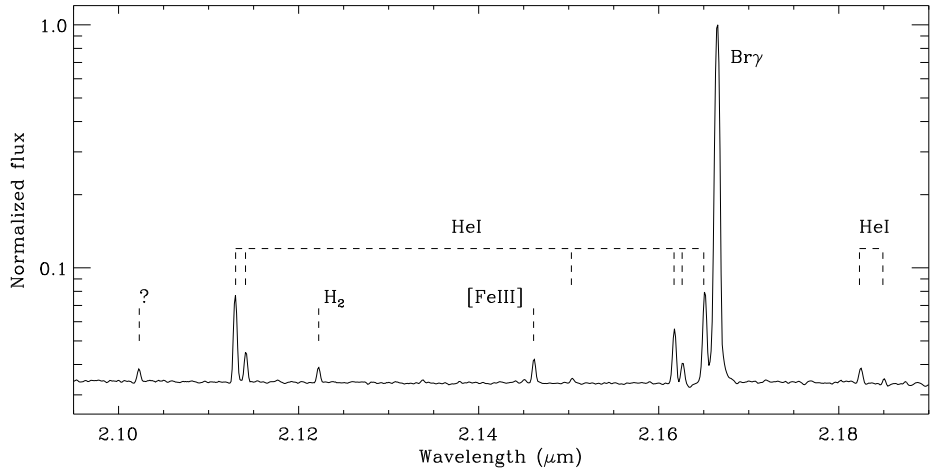}
      \caption{Nebular spectrum of G29.96--0.02 extracted from a 0\farcs3
        $\times$ 0\farcs882 region around the peak of ionized
        emission. The lines observed, their identification and their
        relative flux with respect to Br$\gamma$ are
        listed in Table~\ref{table:vlt:lines}. Note the logarithmic
        scale of the flux scale.}
         \label{fig:vlt:spec}
   \end{figure*}

\begin{table*}[!ht]
\caption{Emission lines in the nebular K-band spectra of G29.96$-$0.02.}
  \label{table:vlt:lines}
  \begin{center}
    \leavevmode
    \normalsize
    \begin{tabular}[h]{lllclll}
      \hline \hline \\[-5pt]
    \multicolumn{1}{c}{$\lambda^\dagger$ (\mum)} &
    \multicolumn{1}{c}{Identification} &
    \multicolumn{1}{c}{Peak flux$^\Diamond$} &&
    \multicolumn{1}{c}{$\lambda^\dagger$ (\mum)} &
    \multicolumn{1}{c}{Identification} &
    \multicolumn{1}{c}{Peak flux$^\Diamond$} 

\\[5pt]  \cline{1-3} \cline {5-7} \\[-5pt]

2.10213$^\flat$ & ?                    & 0.004 &&    
2.16147 & He 7$^3$F--4$^3$D            & 0.024 \\     

2.11282 & He 4$^3$S--3$^3$P            & 0.041 &&    
2.16240 & He 7$^1$F--4$^1$D            & 0.007 \\     

2.11371 & He 4$^1$S--3$^1$P            & 0.011 &&     
2.16495 & He 7$^{1,3}$G--4$^{1,3}$F    & 0.046 \\     

2.12183 & H$_2$ 1--0 S(1)              & 0.005 &&     
2.16612 & H 7--4 (Br$\gamma$)          & 1.00  \\     

2.14570 & \FeIII\ $^3$G$_3$--$^3$H$_4$ & 0.009 &&     
2.18245 & He 7$^3$P--4$^3$D            & 0.005 \\     

2.15007 & He 7$^3$S--4$^3$P            & 0.002 &&     
2.18483 & He 7$^1$D--4$^1$P            & 0.001 

    \\[5pt] \hline

      \end{tabular}
  \end{center}
{\small
($^\dagger$) Vacuum wavelength. The vaccum rest wavelength for the H$_2$
line is taken from \citet{bragg82} and those of the \HeI\ lines from  
\citet{benjamin99}.
($^\Diamond$) Flux with respect to Br$\gamma$ at the position of
the arc peak.
($^\flat$) Observed line center wavelength.}
\end{table*}

   \begin{figure*}[!ht]
   \centering
   \includegraphics[width=11.5cm]{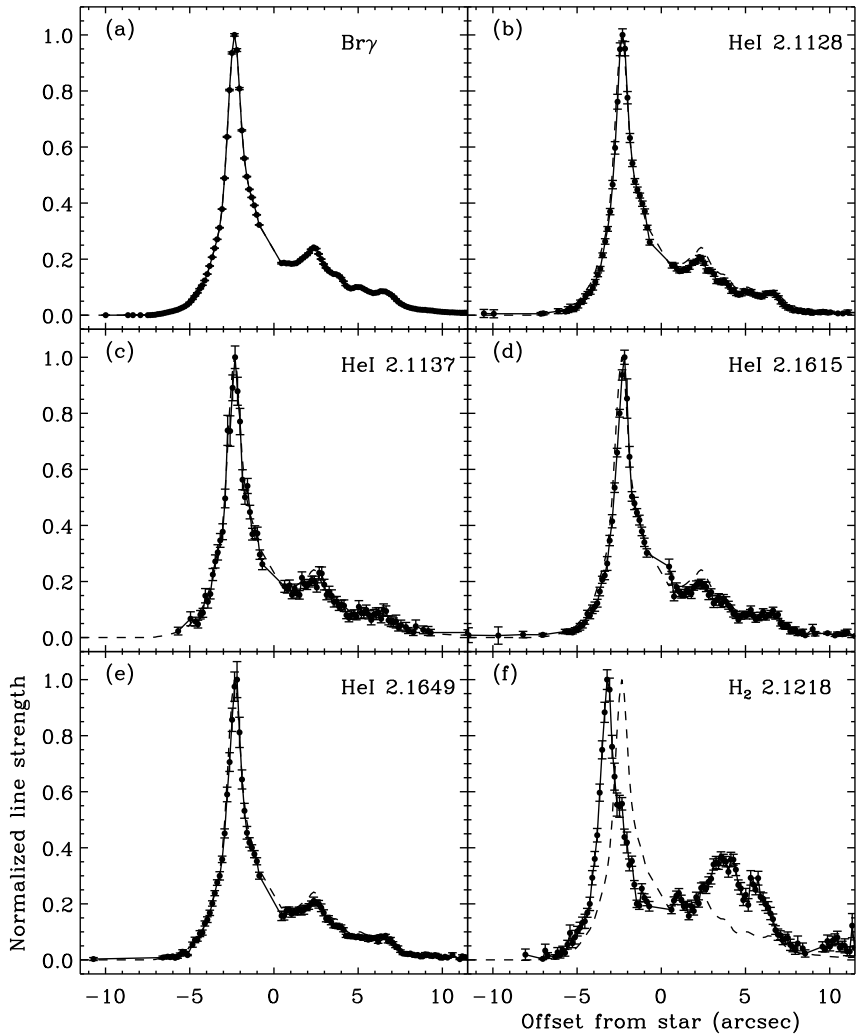}
      \caption{Spatial variation of {\bf a)} Br$\gamma$ , {\bf b--e)}  
      the strongest
      \HeI\ recombination lines  and {\bf f)} H$_2$ 1--0 S(1)
      across the symmetry axis of 
      G29.96 (from left to right in Fig.~\ref{fig:vlt:G29}). 
      The normalized \HeI\ and H$_2$ intensity  distributions are 
      compared to the Br$\gamma$
      profile (dashed line).
      The origin corresponds to the position
      of the ionizing star. At a distance of 6 kpc, 1\arcsec\
      corresponds to 0.029
      pc. While the \HeI\ and \HI\ distributions trace
      each other very well, the H$_2$ line peaks 0\farcs93 ahead
      of Br$\gamma$ and shows a different 
      structure in the tail. We note that it was not possible
      to measure line fluxes close to the position of the
      ionizing star, which explains the absence of points in between 
      $-1$\arcsec\ and 0\arcsec.}
         \label{fig:vlt:profiles}
   \end{figure*}

   \begin{figure*}[!ht]
   \centering
   \includegraphics[width=13.3cm]{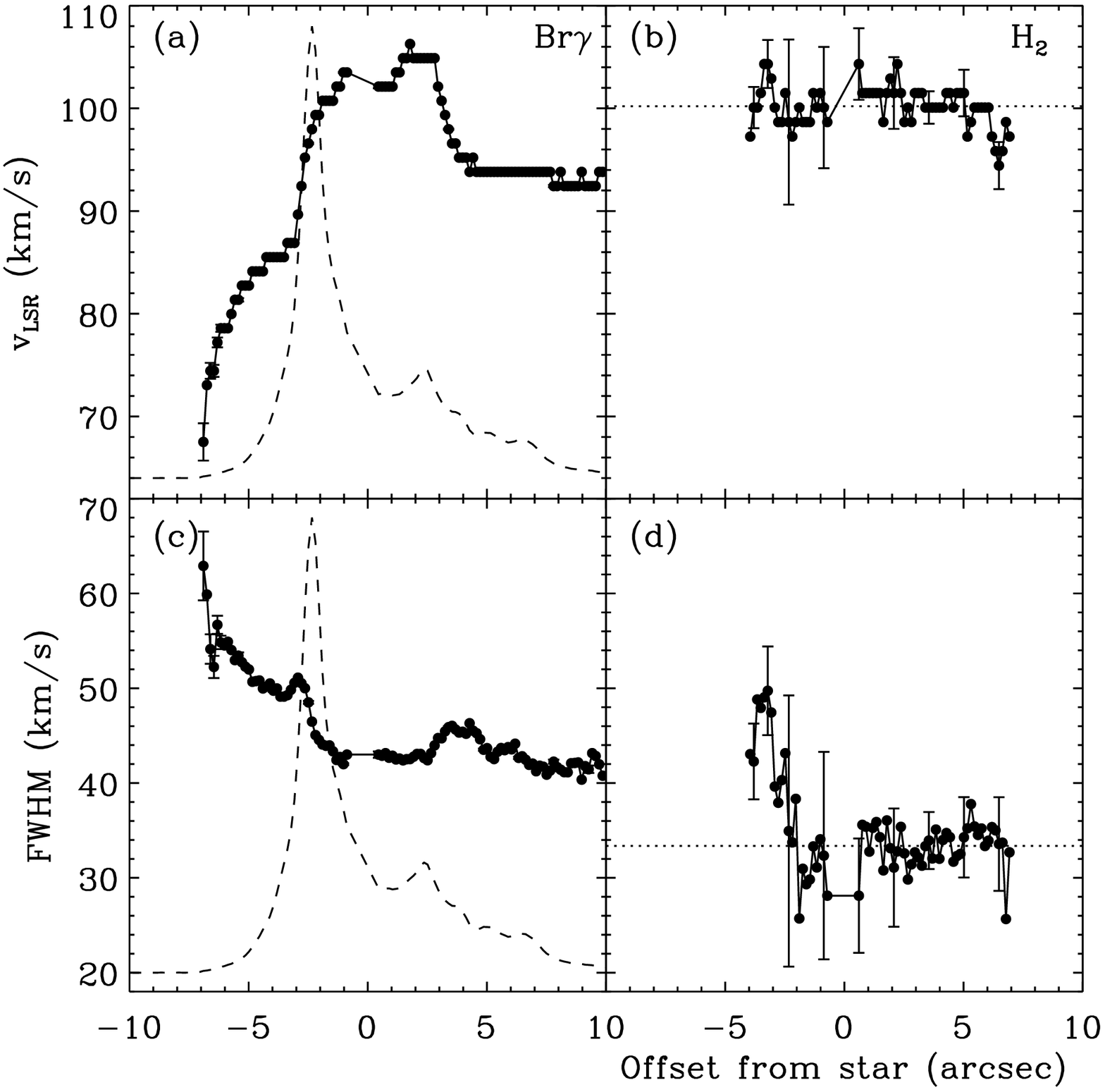}
      \caption{Spatial variation of the velocity centroid (top panel) 
        and line FWHM (bottom
        panel) of
        Br$\gamma$ and H$_2$ along the symmetry axis of G29.96
        (from left to right in Fig.~\ref{fig:vlt:G29}).
        Typical error bars are indicated at different positions along the 
        slit. In panels a) and c),
        the intensity distribution of
        Br$\gamma$ is plotted (dashed line) for comparison. 
        The dotted lines in panels b) and d) represent
        the average H$_2$ centroid velocity and FWHM calculated
        between positions $-3$\arcsec\ and $+5$\arcsec, and
        $0$\arcsec\ and $+5$\arcsec, respectively.      
        The zero point along the slit corresponds to
        the position of the ionizing star. 
        At a distance of 6 kpc, 1\arcsec\ corresponds to 0.029 pc.}
         \label{fig:vlt:vel}
   \end{figure*}

\section{Results}
\label{sect:vlt:results}

Figure~\ref{fig:vlt:spec} shows the nebular spectrum of G29.96 extracted 
from a 0\farcs3 $\times$ 0\farcs882 region around the peak of
the emission. The strongest line corresponds to
Br$\gamma$. The detected \HeI\ emission lines are
the transitions of 4$^3$S--3$^3$P at 2.1128~\mum,
4$^1$S--3$^1$P at 2.1137~\mum, 
7$^3$S--4$^3$P at 2.1501~\mum,
7$^3$F--4$^3$D at 2.1615~\mum,
7$^1$F--4$^1$D at 2.1624~\mum,
the  7$^{1,3}$G--4$^{1,3}$F blend at 2.1649~\mum\ and the
transitions of 
7$^3$P--4$^3$D and 7$^1$D--4$^1$P at 2.1824 and 2.1848~\mum,
respectively. 
From these \HeI\ lines, significant emission along the slit was found
for the transitions of 4$^3$S--3$^3$P at 2.1128~\mum, 4$^1$S--3$^1$P at
2.1137~\mum,  7$^3$F--4$^3$D at 2.1615~\mum\ and the 
7$^{1,3}$G--4$^{1,3}$F blend at 2.1649~\mum.
The 1--0 S(1) transition of H$_2$ 
at 2.1218~\mum\ and the forbidden line of \FeIII\ at 2.1457~\mum\ 
are also evident. 
Table~\ref{table:vlt:lines} lists the observed lines
with their identifications and their fluxes relative to
that of Br$\gamma$ at the peak position.
An unidentified feature is present at 2.1022~\mum. This feature has also
been detected in near-infrared spectra of compact planetary
nebula \citep{lumsden01b}.

We could measure the nebular line emission along $\sim 20$\arcsec\ 
of the slit, 
centered on the position of the ionizing star.
Figure~\ref{fig:vlt:profiles} illustrates
the spatial variation of the strongest lines in this region. 
Moving from right to left along the slit (see Fig.~\ref{fig:vlt:G29}), 
the Br$\gamma$ flux (Fig.~\ref{fig:vlt:profiles}\,a)
slowly increases 
until the position of the star.
After the star, the Br$\gamma$ flux rapidly reaches a maximum and
subsequently decreases.  This maximum is at 2\farcs27 ahead of
the star
and coincides with the arc observed at radio wavelengths 
\citep[e.g.][]{wood89}.
The peak corresponding to the arc is fitted by a Gaussian profile 
with a full width at half maximum
(FWHM) of 1\farcs14.
The global structure of the tail to the right of the star
can be fitted by a semi-Gaussian centered
at the position of the star and with a FWHM of 7\farcs1 and a
height 0.21 times the peak flux.
The following panels (b--f) in Fig.~\ref{fig:vlt:profiles} compare the 
flux distribution of
the strongest \HeI\ lines and the H$_2$ 2.1218 \mum\ line
to that of Br$\gamma$. The \HeI\ and
Br$\gamma$ distributions trace each other very well, indicating that 
the zones of H$^+$
and He$^+$ are of roughly the same size. This indicates that the central
ionizing star must be hot enough to produce ionized helium throughout
the \HII\ region (see Sect.~\ref{sect:vlt:he+} for a detailed discussion).  
The variation in the H$_2$ line strength
resembles the ionized
gas profile well, although it displays some significant deviations.
It peaks 0\farcs93 ahead of the Br$\gamma$ maximum and exhibits a
different structure in the tail of G29.96.
The H$_2$ peak corresponding to the arc 
can be fitted by a Gaussian profile with a FWHM of 
0\farcs87. 
Since the spatial profiles of the \HI, \HeI\ and H$_2$ lines have been
obtained simultaneously in one same setting, the relative separation
of the Br$\gamma$ and H$_2$ peaks is reliable.

The velocity structure of the ionized (as traced by, for instance,
Br$\gamma$) and molecular gas (as traced by H$_2$)
along the symmetry axis of G29.96 is shown in
Fig.~\ref{fig:vlt:vel}. 
Figure~\ref{fig:vlt:vel}\,a plots the centroid
velocity of the Br$\gamma$ line along the central 20\arcsec\ of the
slit. 
There is a large variation in radial velocity ($\sim 38$~km~s$^{-1}$)
across the object. The gas in front of the head of the arc
has a velocity which increases from 67 (position $-7$\arcsec) to 87
\kms\ (position $-3$\arcsec). 
At the arc, between positions $-3$\arcsec\ and $-1$\arcsec,
the velocity increases more rapidly,
reaching a value of 99 \kms\ at the
position of the line intensity peak. It continues to increase towards
the tail where it reaches 105 \kms\ at approximately 3\arcsec\ behind
the star. Afterwards, the velocity drops to a
value of approximately  94 \kms\ at $\sim 5$\arcsec\ behind the star.

Figure~\ref{fig:vlt:vel}\,c shows the behaviour of the Br$\gamma$
FWHM. As can be seen, there is a very
rapid increase of the linewidth ahead of the line intensity
peak, which gets up to $\sim 62$~\kms\ wide. 
The Br$\gamma$ line is $\sim 42-46$~\kms\ wide behind the arc.

This behaviour of the Br$\gamma$ line velocity 
across the symmetry axis of G29.96 agrees well
with the radio recombination line studies of \cite{wood91} and 
\cite{afflerbach94}. 
Low spatial resolution observations of Br$\gamma$ have been presented by
\cite{lumsden96,lumsden99}. In contrast to their results, we see
a larger variation in the peak velocity of the Br$\gamma$ line along
the symmetry axis. In particular, the appreciable increase in velocity
by more than 30 \kms\ just ahead of the arc is not apparent from their
study. This is likely due to the poorer spatial resolution of this earlier
study combined with the steep increase in peak intensity towards the
arc, which gets a high weight in their spatial averaging procedure.

   \begin{figure*}[!ht]
   \centering
   \includegraphics[width=14cm]{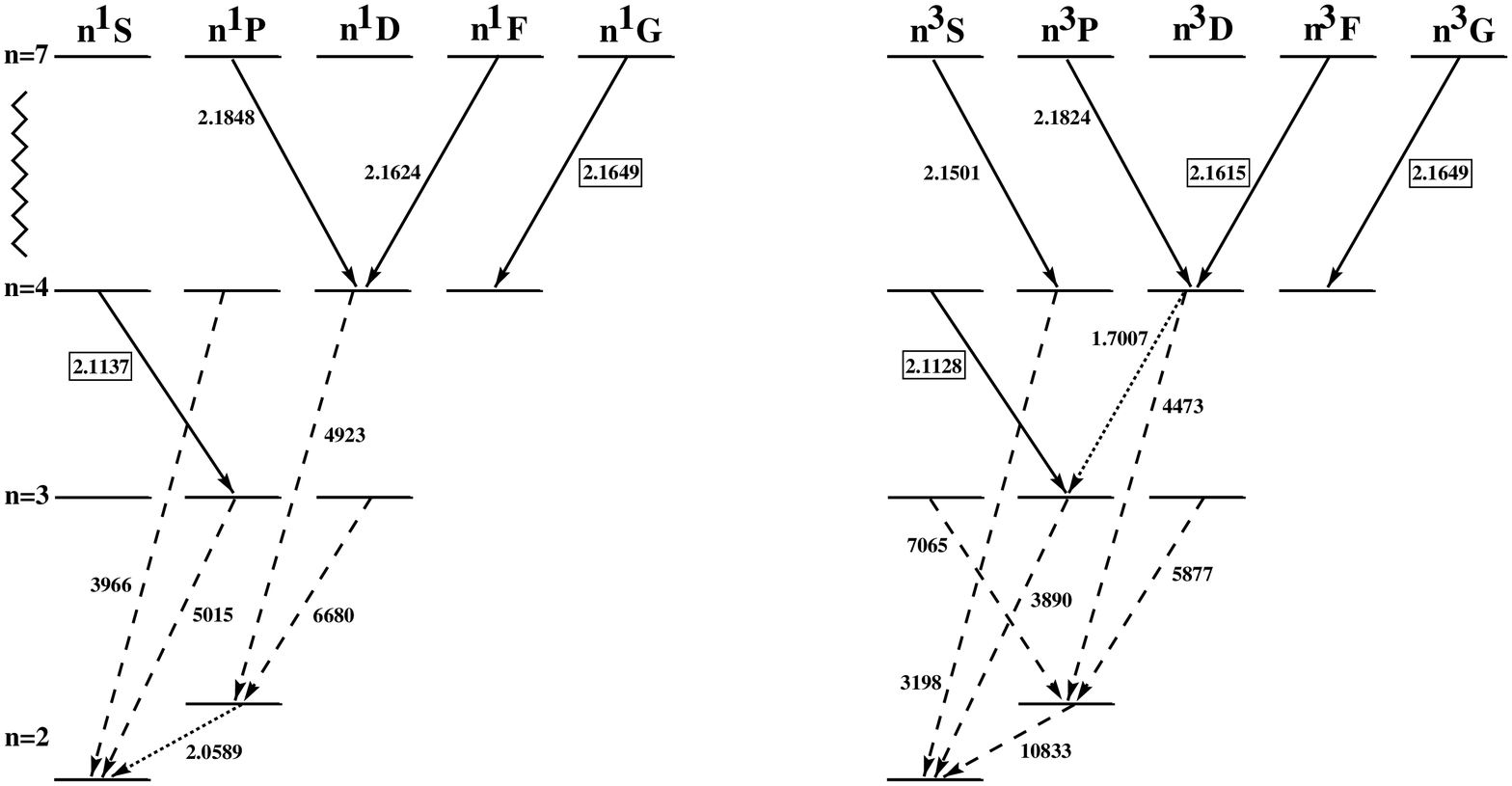}
      \caption{Partial Grotrian energy-level diagram of the \HeI\ singlet and
        triplet transitions  showing the
        infrared lines observed in the K-band spectrum of G29.96 (solid
        arrows), other important infrared lines (dotted  arrows)
        and the strongest optical lines (dashed arrows). Note
        that the energy levels are not drawn to scale. The wavelengths of
        the infrared lines are indicated in \mum; those of the optical
        lines in \AA. The wavelengths of the lines used in the present
        analysis are marked by a box.}
         \label{fig:vlt:he:atom}
   \end{figure*}

The velocity structure observed for Br$\gamma$
is compared to that of the molecular gas traced by the H$_2$ 1--0 S(1)
line (see Figs.~\ref{fig:vlt:vel}\,b and \ref{fig:vlt:vel}\,d). 
There is little variation in 
the peak velocity of the H$_2$ line, which shows an almost constant
value of 100 \kms\ across G29.96
(calculated between positions $-3$\arcsec\ and $+5$\arcsec, where the
H$_2$ line emission is significant, i.e. above 3$\sigma$). 
The average velocity of 100 \kms\ is consistent with the values (around
98 \kms) obtained from single dish observations of rotational transitions
of CS \citep[e.g.][]{churchwell92,bronfman96,olmi99} and
NH$_3$ \citep[e.g.][]{churchwell90b,cesaroni92}. VLA measurements of
ammonia by \cite{cesaroni94} resulted in a value of 98.7~\kms\  
towards the hot core detected just in front of the arc of G29.96.
Thus, the H$_2$ gas is practically at rest with respect to the ambient medium.
The H$_2$ FWHM shows a practically constant value of 33 \kms\ in the
interval [0\arcsec, $+$7\arcsec]. This FWHM is of the order of the
spectral resolution, implying that the H$_2$ line is
unresolved between these positions. Similarly to the case of the
ionized gas, a significant increase of the FWHM 
up to $\sim 50$~\kms\ is observed in front of the star.

The K-band spectral observations of G29.96 by \cite{lumsden99},
although of much lower spatial resolution than ours, have about twice
our spectral resolution ($\sim 18$~\kms). Their observations of the $1-0$ S(1)
H$_2$ line show, as well, a very
rapid increase in linewidth (from $\sim 20$ to 28 \kms\ in their case) 
ahead of
the arc. They were also able to discern small variations of the
H$_2$ centroid velocity in the same direction as those of the
ionized gas. This behaviour is, however,  not evident from our data.

Comparing the Br$\gamma$ and H$_2$ velocity structure, we note that 
around the position of the
ionizing star (between positions $-2$\arcsec\ and $+3$\arcsec), 
the ionized gas is systematically
redshifted relative to the molecular gas by a few (2--5)~\kms, 
whereas in the tail and ahead of
the arc it becomes progressively more blueshifted.     
The increasing blueshift of the ionized gas in front of the arc
coincides with the significant increase in the Br$\gamma$ and H$_2$ FWHM. 
This pattern has to be due to acceleration of the ionized gas along the
line of sight. 

\section{The recombination lines}
\label{sect:vlt:theory}

The nebular spectrum of G29.96 is dominated by recombination lines of hydrogen
and helium. For most of the \HI\ emission lines observed in nebulae,
there are no radiative transfer problems; hence, the Case B approximation 
\citep{hummer87},
which assumes that the Lyman transitions are optically thick,
reproduces well, apart from dust extinction, 
the observed nebular spectra. The recombination
radiation of \HeI\ singlets is very similar to that of \HI, and can also
be approximated by the Case B treatment. In this case, by considering
Case B it is assumed that the nebula has a large optical depth in
transitions arising from the  1$^1$S level. However, the recombination
cascade of the \HeI\ triplets can be  modified by the fact that the
2$^3$S level (see Fig.~\ref{fig:vlt:he:atom}) is highly metastable. 
All recaptures to triplets eventually
cascade down to the 2$^3$S level. Depopulation of this level can
only occur through photoionization, through 
collisional transitions to 2$^1$S and 2$^1$P, or through the forbidden 
2$^3$S--1$^1$S transition. The rates of these processes are small and
hence, the population of the 2$^3$S level is large; in turn, the optical depth
of the lower n$^3$P--2$^3$S transitions is substantial. Moreover, the
collisional transitions from 2$^3$S to upper levels become significant
and the emission from these levels is enhanced with respect to the
predictions of a pure recombination model. Because of this
interplay of collisional and radiative transfer effects, the triplet
lines become sensitive to the density and optical depth.

Recently, \cite{benjamin02} have considered the effect of the optical
depth of the 2$^3$S level on the recombination spectrum of a
spherically symmetric nebula. They used a model atom with individual
levels up to $n=20$ and parameterized their calculations in
terms of the line center optical depth of the 3$^3$P--2$^3$S 3890 \AA\
line. A FORTRAN program
to calculate emissivities of lines arising from levels with $n \leq
10$ over a large range of nebular conditions is available from 
\citeauthor{benjamin02}\footnote{http://wisp.physics.wisc.edu/$\sim$benjamin} 

   \begin{figure}[!ht]
   \centering
   \includegraphics[width=8cm]{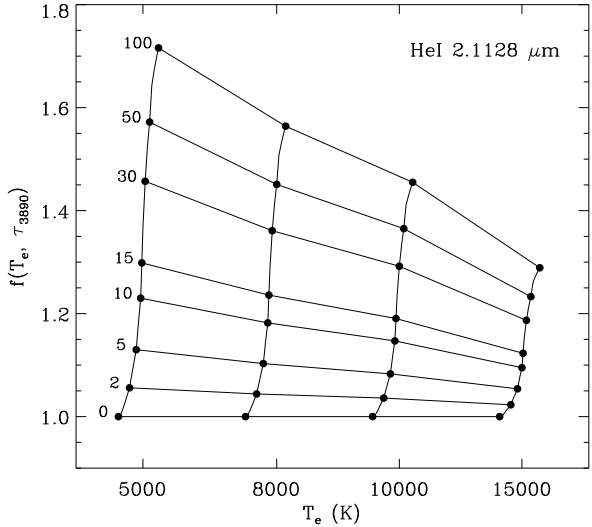}
      \caption{Optical depth correction factor, $f$, for the 2.1128 \mum\
        \HeI\ line as a function of the electron temperature for
        $\tau_{3890}$ between 0 and 100 and an electron density
        \nel~=~$10^4$~\cm3.}
         \label{fig:vlt:tau}
   \end{figure}

We use this program to estimate the Case B line fluxes of the \HeI\ lines
observed in the spectrum of G29.96 and their possible enhancement due
to collisional excitation from the 2$^3$S level, or as a result of
significant opacity in the n$^3$P--2$^3$S series.

Although \cite{benjamin02} include the collisional coupling between
the triplet and singlet levels, they find that only the
triplet lines arising from low levels show a noticeable enhancement 
in strength with respect to the zero
optical depth case. From the triplet lines observed in the spectrum of
G29.96, we find that the effect of the finite optical depth of the  2$^3$S level
is only evident in the transition of 4$^3$S--3$^3$P at 2.1128
\mum. Adopting the nomenclature by \cite{benjamin02}, we define the
optical depth correction factor 
$f_{\rm line}$(\nel,\Tel,$\tau_{3890}$)=
$j$(\nel,\Tel,$\tau_{3890}$)/$j$(\nel,\Tel,0), i.e. the
ratio of the line emissivity for optical depth  $\tau_{3890}$ to the
emissivity for zero optical depth. Figure~\ref{fig:vlt:tau} presents the
variation of
the optical depth correction factor for the 2.1128 \mum\ line as a function
of $\tau_{3890}$ and \Tel\ for the case of a nebula with \nel=$10^4$
\cm3. 

The optical depth of the 3$^3$P--2$^3$S 3890 \AA\ line in G29.96 can
be roughly estimated. We have that 
$\tau_{3890}=n(2^3{\rm S}) \sigma_{3890} R$, where 
$n(2^3{\rm S})$ is the density of atoms in the metastable state, 
$\sigma_{3890}$ is the cross section at the line center of the 
3890 \AA\ line and $R$ is the radius of the nebula. 
The relative population of the $2^3{\rm S}$ level depends on the local
electron density and temperature, and can be estimated using equation
5 in \cite{kingdon95}. Assuming an electron density of 2$\times10^4$
\cm3\ and an electron temperature of 6500 K \citep{afflerbach94}, we
find  that $n(2^3{\rm S})/n({\rm He}^+)\sim 8 \times 10^{-6}$ in the
tail of G29.96.
Equation~1 
in \cite{benjamin02} gives that $\sigma_{3890} \sim 7 \times 10^{-14}$
cm$^2$ for \Tel $=6500$ K and a turbulent velocity of 0 km s$^{-1}$. Assuming
a helium abundance of 10\% by number and a nebular radius of 7\arcsec\
(see Sect.~\ref{sect:vlt:results}),
which at a distance of 6 kpc corresponds to $\sim 0.20$ pc, we
obtain $\tau_{3890} \gg 100$. Considering a turbulent velocity
of 20~\kms\ and a filling factor of 0.1, we obtain $\tau_{3890} \sim 80$.
Hence, the  2.1128~\mum\ line is
expected to be enhanced 
(see Fig.~\ref{fig:vlt:tau}) with respect to the pure recombination model.

It is illustrative at this point to use the relative strengths of the \HeI\
recombination lines to (1) determine how closely these lines follow
the recombination theory and (2) study their dependence on the local
nebular conditions. The ratio between any two \HeI\ lines is
insensitive to the electron density, but can 
depend on the electron temperature.

\subsection{The electron temperature}
\label{sect:vlt:te}

   \begin{figure*}[!ht]
   \centering
   \includegraphics[width=18cm]{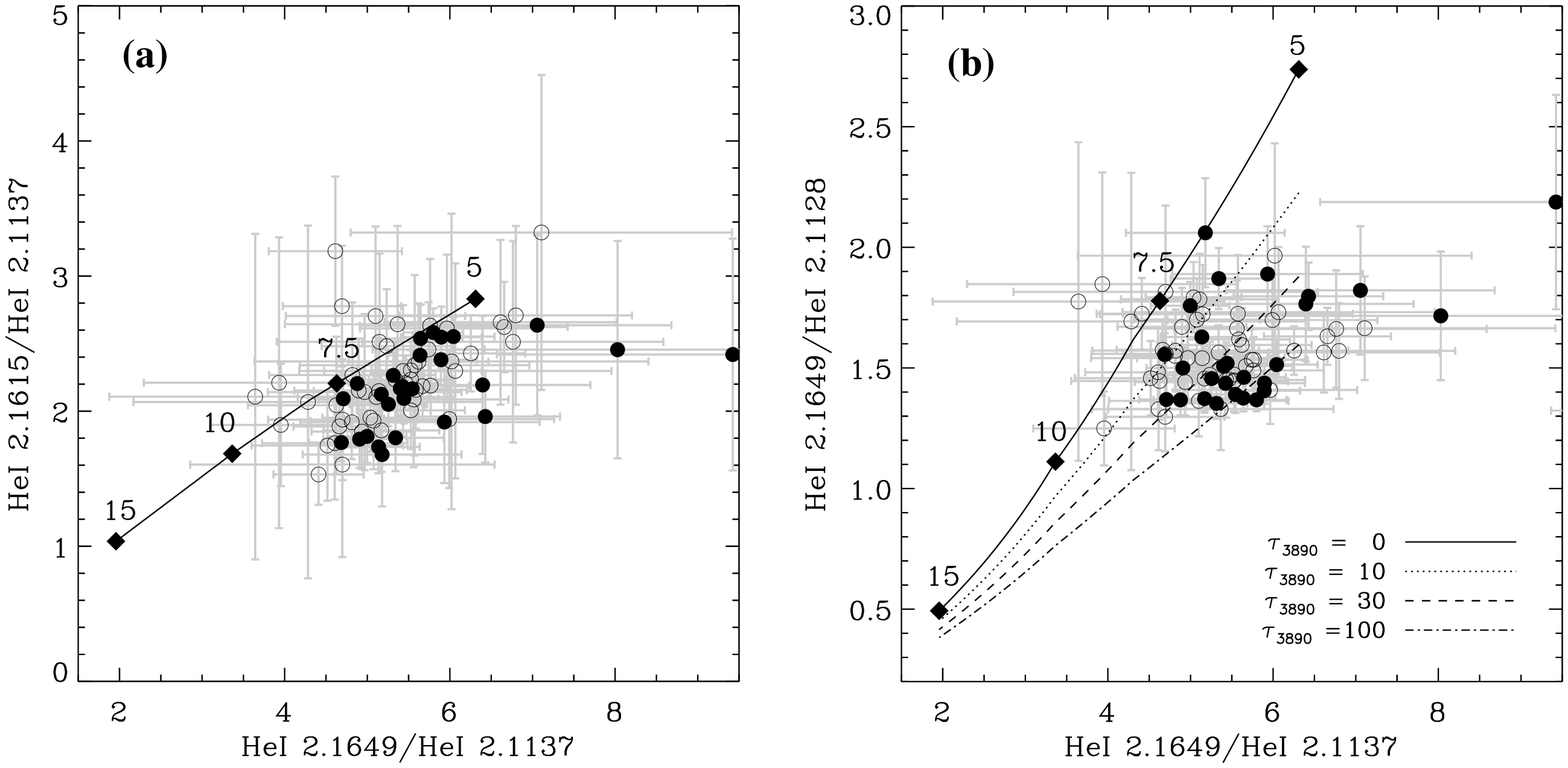}
      \caption{{\bf a)} Plot of
        the line ratio 2.1615-7$^3$F--4$^3$D/2.1137-4$^1$S--3$^1$P 
        against the ratio 
        2.1649-7$^{1,3}$G--4$^{1,3}$F/2.1137-4$^1$S--3$^1$P. 
        Note that these line ratios should not depend on radiative
        transfer effects. 
        {\bf b)} 
        2.1649-7$^{1,3}$G--4$^{1,3}$F/2.1128-4$^3$S--3$^3$P
        against 2.1649-7$^{1,3}$G--4$^{1,3}$F/2.1138-4$^1$S--3$^1$P.
        In both figures, the solid line represents
        the Case B predictions for \Tel\ = 5, 7.5, 10 and 15 $\times 10^3$
        K (indicated by solid diamonds) and \nel\ $=10^4$ \cm3.
        The data indicate a temperature  between 
        5000 and 7500~K.
        The non-solid lines in b) represent the calculations by
        \cite{benjamin02} for non-zero optical depth of the 2$^3$S
        metastable level.
        Only points located in the
        interval [$-5$\arcsec, $+8$\arcsec] are considered.
        The data points from the tail of G29.96 ($> 0$\arcsec) are
        plotted as open circles; those from the arc ($< 0$\arcsec) 
        are plotted as solid circles. 
        The open and filled circles have very similar distributions.} 
         \label{fig:vlt:color}
   \end{figure*}

Figure~\ref{fig:vlt:color}\,a shows the relation
between the \HeI\ line ratios
2.1615-7$^3$F--4$^3$D/2.1137-4$^1$S--3$^1$P 
and 2.1649-7$^{1,3}$G--4$^{1,3}$F/2.1137-4$^1$S--3$^1$P. These ratios
show the largest dependence on \Tel. 
The calculations by \cite{benjamin02} show that these ratios are 
expected to be independent of radiative transfer problems
arising from the metastable 2$^3$S level.
The solid line represents the theoretical predictions by
\cite{benjamin02} for \Tel\ = 5, 7.5, 10, 15~$\times 10^3$~K and 
\nel\ $=10^4$ \cm3. The model reproduces the trend in the 
data reasonably well, although
there is a slight mismatch between
data and theory. The data points are, on average, 20\% below the model
predictions. 
The origin of this discrepancy is unclear. 
It is clear from Fig.~\ref{fig:vlt:color}\,a
that the electron temperature of the ionized gas is in between 
5000 and 7500 K, and that both the points from the tail of G29.96 ($>
0$\arcsec) and those from the arc ($< 0$\arcsec) are distributed equally.
 
Figure~\ref{fig:vlt:color}\,b shows the relation between the  
\HeI\ line ratios 2.1649-7$^{1,3}$G--4$^{1,3}$F/2.1128-4$^3$S--3$^3$P 
and 2.1649-7$^{1,3}$G--4$^{1,3}$F/2.1137-4$^1$S--3$^1$P. The data show
a significant spread in these line ratios, largely exceeding the
quoted uncertainties. We conclude that the 2.1649/2.1128~\mum\ line ratio is
sensitive to the optical depth and that the 2.1128~\mum\ line is
enhanced with respect to the pure recombination case (solid line).
A comparison with the model calculations \citep{benjamin02}
indicates optical depths, $\tau_{3890}$, in between 0 and 100, as
expected (see above).

   \begin{figure}[!ht]
   \centering
   \includegraphics[width=8.7cm]{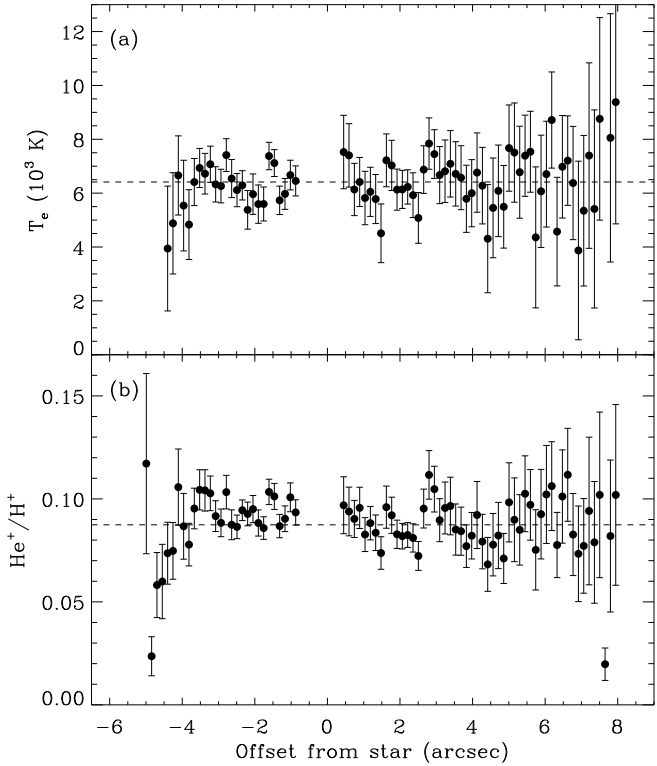}
      \caption{{\bf a)} Electron temperature structure along the symmetry
       axis of G29.96 (from left to right in Fig.~\ref{fig:vlt:G29}) as
       derived from the \HeI\ line ratio 2.1649/2.1138 \mum. The dashed
       line represents the average \Tel\ of 6400~K.
       {\bf b)} He$^+$/H$^+$ structure. The dashed line represents
       the average value of 0.087.
       Only points located in the
       interval [$-5$\arcsec, $+8$\arcsec] are considered.} 
       \label{fig:vlt:prop}
   \end{figure}

A detailed electron temperature profile along
the axis of G29.96 can be obtained by using the ratio of any two of
the \HeI\ recombination lines. Figure~\ref{fig:vlt:color}\,a shows that the
2.1649/2.1137~\mum\ and 2.1615/2.1137 \mum\ line ratios are reasonably well
reproduced by the Case B approximation and can be used to probe 
variations in \Tel. 
Hence, by comparing the observed line ratios at any position with the Case
B predictions we can measure the variation of \Tel\ along the
slit.
 
Figure~\ref{fig:vlt:prop}\,a shows the \Tel\ profile along the symmetry
axis of G29.96
as derived from the 2.1649/2.1137 \mum\ line ratio. We see that the
electron temperature is, within the uncertainties, constant along the
slit with a mean value of 6400 $\pm$ 150 K. When we use the 
2.1615/2.1137~\mum\ line ratio, we also obtain a
similarly constant \Tel\ structure along the slit, but with a higher
average value of 7550 $\pm$ 150 K. The difference between these
two determinations comes from the discrepancy between data and
theory shown in Fig.~\ref{fig:vlt:color}\,a. 

Our measurement of \Tel\ is in good agreement with previous
determinations.
The non-LTE analysis of radio recombination line observations by 
\cite{afflerbach94} resulted in electron temperatures in between 6200 and
8600 K. They estimate an average \Tel\ of
$\sim 6500$~K. The electron temperatures derived by \cite{wood91}
from H76$\alpha$ recombination line observations towards G29.96 are
low, ranging 
from $\sim 2500$~K in the tail to 4200~K in the arc. However,
for this line, deviations from LTE 
can lead to  temperature estimates lower than the true
nebular temperature. Their best estimate of the electron temperature
using non-LTE models is $\sim 5000$~K throughout the
nebula. Finally, \cite{watson97a}  obtained temperatures in between 5000 and
7000 K over the nebula using 2, 6 and 21~cm continuum data. 

Our estimate of \Tel\ is also consistent with the Galactocentric
distance of G29.96.
The electron temperature in \HII\ regions has been shown to increase
with the distance to the Galactic Center as a consequence of the
decrease in metal content. This \Tel\ gradient is approximately given by 
\Tel=5000+5000$\times$\rgal/15 \citep{shaver83,afflerbach96,deharveng00}.
At the Galactocentric distance of G29.96 (4.6 kpc), the expected \Tel\ is
6500 K, consistent with our value. This electron temperature corresponds
to an enhancement of the metal content by a factor of 
$\sim 2$ with respect to that in the solar neighbourhood. This
metallicity is in agreement with the analysis based on the infrared
fine-structure lines observed towards G29.96 
\citep[see][]{martin:paperii}.

\subsection{He$^+$/H$^+$} 
\label{sect:vlt:he+}

The He$^+$/H$^+$ structure along the symmetry axis of G29.96 can be determined
from the ratio of a singlet \HeI\ line to Br$\gamma$ once
a specific model for the \Tel\ structure of the \HII\ region
has been assumed (such a ratio is insensitive to the electron
density). 
The only singlet line we were able to measure along the slit is 
the transition of 4$^1$S--3$^1$P at 2.1137 \mum.

\begin{figure}[!ht]
   \centering
   \includegraphics[width=8.7cm]{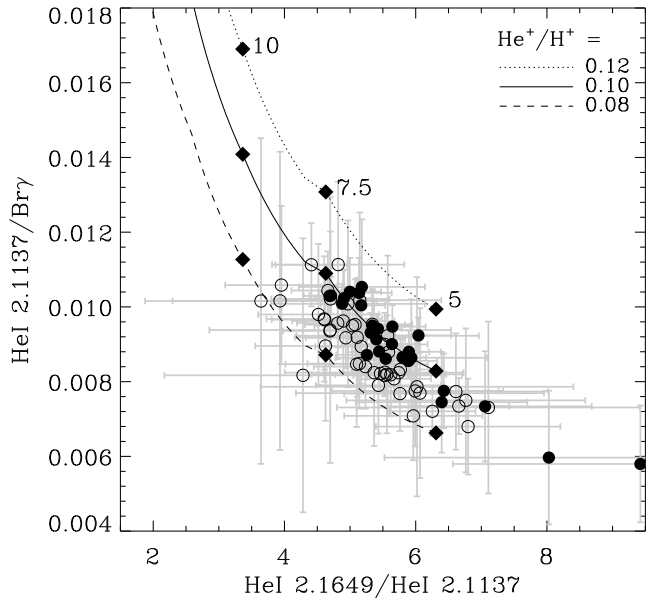}
      \caption{The line ratio \HeI\ 2.1649/Br$\gamma$ is plotted against 
        the ratio \HeI\ 2.1649/2.1137. The lines
        represent the theoretical ratios for 3 different He$^+$/H$^+$
        abundances (0.08, 0.10 and 0.12) and an electron 
        density of $10^4$ \cm3. The values
        corresponding to \Tel~=~5, 7.5 and 10 $\times 10^3$ K are
        indicated by solid diamonds.
        As in previous figures, 
        only points located in the
        interval [$-5$\arcsec,~$+8$\arcsec] are considered, 
        the data points in the tail of
        G29.96 are plotted as open circles and those from the arc as
        solid circles.}
         \label{fig:vlt:heioniz_te}
   \end{figure}

Figure~\ref{fig:vlt:heioniz_te} shows the relation between this
ratio and the \Tel-indicator \HeI\ 2.1649/2.1137 \mum.
There is an (anti)correlation between these two line ratios
because of the different \Tel\ dependence for the \HeI\
single/triplet and the \HI\ lines.
The observed data points are well matched by He$^+$/H$^+$
abundances in between 0.08 and 0.10 and electron temperatures in
between 5000 and 7500 K. A preponderance of low He$^+$/H$^+$
abundances in the tail region is evident from this figure. This may
reflect a He$^+$ zone somewhat smaller than the H$^+$ zone. Such a
difference would be spatially more prominent in the low density tail.

The detailed He$^+$/H$^+$
structure along the symmetry axis of G29.96, calculated for 
\Tel~=~7000~K, is plotted in Fig.~\ref{fig:vlt:prop}\,b. 
We find that He$^+$/H$^+$ is practically constant
across the slit.
We obtain an average  He$^+$/H$^+$
abundance of $0.087 \pm 0.001$.  
Considering a total helium abundance by number
of $\sim 0.1$, this implies that He is practically singly ionized throughout
the nebula.

\cite{kim01} estimated the value of He$^+$/H$^+$ at different
positions in G29.96
from the ratio of the ratio of the radio recombination lines He76$\alpha$ and 
H76$\alpha$. They derived values in between 0.063 and 0.080, slightly
lower than our estimate of 0.087. Their lower estimate could be
due, however, to non-LTE effects on the radio recombination lines that
may not have been taken into account, or to beam smearing.

   \begin{figure}[!ht]
   \centering
   \includegraphics[width=8.7cm]{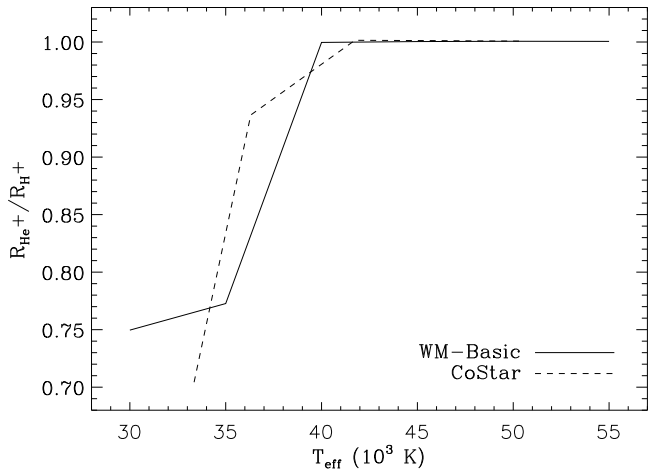}
   \caption{The relative extent of the He$^+$ region to that of 
     the H$^+$ region, $R_{\rm He^+}/R_{\rm H^+}$, is plotted against
     $T_{\rm eff}$, the effective temperature of the ionizing star.
     For $T_{\rm eff} \geq 40\,000 $ K, the He$^+$ and
     H$^+$ zones are coincident.}
         \label{fig:vlt:cloudy}
   \end{figure}

We have already mentioned in Sect.~\ref{sect:vlt:results} that the
central ionizing star in G29.96 must be hot enough to produce ionized
helium throughout the \HII\ region. In order to investigate the
dependence of the relative extent of
the He$^+$ region compared to that of the H$^+$ region, 
$R_{\rm He^+}/R_{\rm H^+}$, 
on the effective
temperature of the star, we have calculated
a set of nebular models with the photoionization code CLOUDY version 94.00
(\citeauthor{ferland00}
\citeyear{ferland00}\footnote{see also http://thunder.pa.uky.edu/cloudy}) 
using MICE, the IDL interface for CLOUDY created by
H. Spoon\footnote{http://www.astro.rug.nl/$\sim$spoon/mice.html}. We
compute nebular models for a static, spherically symmetric,
homogeneous gas distribution with one ionizing star in the center. An
inner cavity with a radius equal to 10$^{17}$~cm is set. Two different
sets of stellar atmosphere  models are taken to describe the spectral 
energy distribution
(SED): the CoStar models by \cite{schaerer97} and the WM-Basic models by
\cite{pauldrach01}. We use stellar models for main sequence (dwarf)
stars. The
number of hydrogen ionizing photons emitted by the central source is
fixed to that of the SED used, the metallicity of the nebula is set
to twice solar and the density equal to 10$^4$ \cm3, appropriate for
G29.96 \citep{afflerbach94,wood91}.

Figure~\ref{fig:vlt:cloudy} shows $R_{\rm He^+}/R_{\rm H^+}$ as a function
of the effective temperature of the star, $T_{\rm eff}$. It is seen
that for $T_{\rm eff} \geq 40\,000 $ K, the He$^+$ and H$^+$ zones are
coincident, and that this result is independent of the SED used. Using
the calibration by \cite{vacca96}, this corresponds to a spectral type
earlier than O7.5. This calibration is, however, based on
plane-parallel models which do not incorporate stellar winds and line
blocking/blanketing. A more recent calibration by \cite{martins02},
based on non-LTE line blanketed atmosphere models with stellar winds
computed with the {\sc cmfgen} code of \cite{hillier98}, yields a spectral
type equal to or earlier than O6. This spectral type determination is
compared to estimates using other methods in Sect.~\ref{sect:vlt:sptype}.

\section{The ionizing star of G29.96}
\label{sect:vlt:sptype}

The opening up of the infrared window has
favoured in recent years the use of infrared spectra to infer the
spectral type of the ionizing stars in heavily extincted regions. The
first spectral classification in the K-band of a star ionizing an 
UCH{\sc ii} region was that of G29.96 \citep{watson97b}.  
The K-band classification \citep{hanson96} is based on the presence 
and strength of
photospheric lines of \HeI\ (2.058 and 2.1126 \mum), 
\ion{C}{iv} (2.078 \mum), \ion{N}{iii} (2.1155 \mum),
\HI\ (2.1661 \mum) and \ion{He}{ii} (2.1885 \mum). 
The \HI\ and \HeI\ lines, however, are heavily contaminated by nebular
emission. Based on the other features, the
spectral types that can be determined from the direct observation of
the K-band spectrum are: O3--O4 (\ion{N}{iii}~ emission  and 
\ion{He}{ii}~ absorption, but no \ion{C}{iv}~ emission), O5--O6.5 (both  
\ion{N}{iii}~ and \ion{C}{iv}~ emission and \ion{He}{ii}~ absorption), 
O7--O8 (\ion{N}{iii}~ emission  and \ion{He}{ii}~ absorption, and weak 
\ion{C}{iv}~ emission), and O9 or later (none of these). Using this
classification scheme, \cite{watson97b} restricted the ionizing star in
G29.96 to spectral type O5--O6.5.

\begin{figure}[!ht]
   \centering
   \includegraphics[height=8.8cm,angle=90]{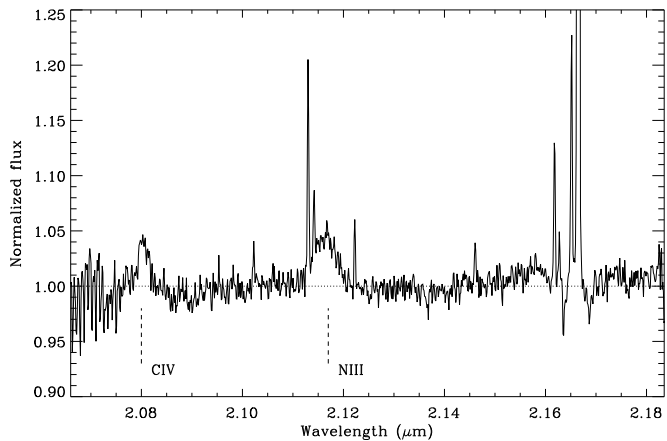}
      \caption{Spectrum of the ionizing star in G29.96. The emission
        features of \ion{C}{iv} at 2.080 \mum\ and \ion{N}{iii} at
        2.116 \mum\ are indicated. These features constitute the base of the
        K band classification. The lines of \HI\ and \HeI\ are
        contaminated by nebular emission.} 
         \label{fig:vlt:star}
   \end{figure}

Figure~\ref{fig:vlt:star} shows the spectrum of the ionizing star of
G29.96 extracted from the slit (\citeauthor{kaper02a}
\citeyear{kaper02a}\footnote{We note that Kaper et al. wrongly
  identify the \HeI\ 2.1649~\mum\ line as \HeII\ 2.1652~\mum, and use
  this identification to argue that the ionizing star is of even
  earlier spectral type than O5.}; Bik et al. in prep.). 
The emission features of \ion{C}{iv}~
and \ion{N}{iii}~ are clearly present. The spectral coverage of our
observations was not large enough to include the \ion{He}{ii}~
line at 2.1885 \mum. The presence and strength of the  \ion{C}{iv}~
and \ion{N}{iii}~ features is consistent with the estimate by
\cite{watson97b}, i.e. with an O5--O6.5 star. \cite{watson97b} note
that spectral types as late as O7--O8 could also be consistent with
the data  if the \ion{C}{iv}~ and \ion{N}{iii}~ features were enhanced
as a consequence of G29.96 being a region with a metallicity twice
that of the Sun. However, these late spectral types are ruled out
because they are not hot enough to produce a He$^+$ zone coincident
with that of  H$^+$ (see Sect.~\ref{sect:vlt:he+}). Thus, the K-band
spectrum of the ionizing star, together with the condition imposed by
the extent of the He$^+$ zone (equal to or earlier than O6), 
limits the spectral type to O5--O6.

\begin{table*}[!ht]
\caption{Different estimates of the spectral type of the ionizing 
  star in G29.96.}
  \label{table:vlt:sptype}
  \begin{center}
    \leavevmode
    \normalsize
    \begin{tabular}[h]{llll}
      \hline \hline \\[-5pt]
    \multicolumn{1}{c}{Method} &
    \multicolumn{1}{c}{\Teff\ (kK)} &
    \multicolumn{1}{c}{Spectral type$^\triangle$} &
    \multicolumn{1}{c}{Reference} \\[5pt] \hline \\[-5pt]

Bolometric luminosity & $< 49^\dagger$ & $<$ O3$^\star$ & 1, 2\\

Lyman continuum photon flux & $ 35-40^\dagger$ 
& O8 -- O6$^\star$ & 3, 4, 5, 6, 7\\

Near-IR photometry & $\leq 42.5$ & $\leq$ O5$^\star$ & 2\\

Nebular IR lines & 32 -- 38 & O9 -- O6.5$^\star$ & 8\\

Extent of the He$^+$ zone & $\geq 40^\sharp$
& $\geq$ O6$^\sharp$ & 9\\

K-band spectrum & 38 -- 42.5$^\star$ & O6.5 -- O5 & 9,10

\\[5pt] \hline

      \end{tabular}
  \end{center}
{\small
($^\triangle$) ``$>$'' means ``earlier than'', and ``$<$'' means 
``later than''.
($^\dagger$) Adopting the calibration by \cite{vacca96}.
($^\star$) Adopting the \Teff\ scale by \cite{martins02} for dwarf stars.
($^\sharp$) See Sect.~\ref{sect:vlt:he+}.
References: 
 (1) \cite{peeters:catalogue}; 
 (2) \cite{watson97a};
 (3) \cite{wood89}; 
 (4) \cite{cesaroni94}; 
 (5) \cite{becker94};
 (6) \cite{fey95};
 (7) \cite{kim01};
 (8) \cite{morisset:paperiii};
 (9) this work;
(10) \cite{watson97b};
}
\end{table*}


Various other estimates of the spectral type of the dominant ionizing star
of G29.96 have been obtained from very different observations and
methods. We briefly discuss these estimates using up-to-date stellar
models and homogeneous assumptions. In particular, all observational
constraints are scaled to a distance of 6 kpc and we use the latest
temperature scale of O stars by \cite{martins02}.
We assume that the ionizing star is an unevolved zero age
main sequence (ZAMS) star, although we note that some authors
\citep{watson97a,morisset:paperiii} suggest that it may be a more
evolved star.
These estimates are (see also Table~\ref{table:vlt:sptype}): 

\begin{itemize}

\item[$\bullet$] {\it Bolometric luminosity} The total bolometric luminosity
obtained from the IRAS fluxes \citep{peeters:catalogue} and the
overall SED \citep{watson97a} in an arcminute-sized region 
is log$(L/L_{\sun})=5.9$. Since this luminosity is integrated over a
large region, it is likely to include contributions from other
sources, for instance, the hot core. Hence, it must be considered as
an upper limit to the bolometric luminosity of the ionizing star.
This upper limit to the luminosity corresponds to 
a star with an effective temperature lower than  49\,000~K
\citep{vacca96}. Adopting the
\Teff-spectral type calibration by \cite{martins02} for dwarf stars, this
corresponds to a star later than O3.

\item[$\bullet$] {\it Number of Lyman continuum photons} The number of Lyman
continuum photons derived for G29.96 from radio continuum observations
is in the range log$(N_{\rm Lyc})=48.43-48.93$ s$^{-1}$
\citep{wood89,cesaroni94,becker94,fey95,kim01}, which corresponds
to an effective temperature in between 35\,000 and 40\,000~K \citep{vacca96}. 
This \Teff\ range  corresponds to a star with a spectral
type of O8--O6 \citep{martins02}. 

\item[$\bullet$] {\it Near-infrared photometry} Constraints on the 
K magnitude and
the bolometric luminosity allowed \cite{watson97a} to place limits on
the location of the ionizing star in the H-R diagram. They found a
3$\sigma$ upper limit on the effective temperature of 42\,500~K. 
Adopting the \Teff-spectral type calibration by \cite{martins02},
this corresponds to a star equal to or later than O5.

\item[$\bullet$] {\it Nebular infrared lines} The ratios of nebular
fine-structure lines  (\ArIII/\ArII\ 9.0/7.0, 
\SIV/\SIII\ 10.5/18.7, \NeIII/\NeII 15.5/12.8 and \NIII/\NII\ 57/128
\mum) observed by ISO \citep{peeters:catalogue}
have been used by \cite{morisset:paperiii} to 
constraint the properties of the ionizing star in G29.96. They use the
most recent non-LTE line blanketed atmospheres with stellar winds and
obtain effective temperatures in the range 32\,000--38\,000 K. 
Adopting the \Teff-spectral type calibration by \cite{martins02},
this corresponds to a star with a spectral type O6.5--O9.

\end{itemize}

Practically all the indirect methods to derive the spectral type which 
have been listed above agree well with the spectral type derived from
the K-band spectrum. The exception is the method based on the nebular
infrared lines, which predicts a slightly later type. In 
\cite{martin:metal} it
was indicated that the infrared fine-structure line ratios such as
\NeIII/\NeII\ 15.5/12.8~\mum\ are influenced by the stellar and nebular
metallicity. Hence, their diagnostic use as indicator of the degree of
ionization of the nebula is conditioned to an adequate treatment of
the effect of metallicity on the stellar UV spectrum, i.e, on the line
blocking/blanketing and stellar wind.

\section{The dynamics of G29.96}
\label{sect:vlt:dynamics}

Any model for the formation and evolution of the UCH{\sc ii} region
G29.96 should be able to explain the following characteristics:

\begin{enumerate}

\item The general morphology of the nebula is described by a bright arc 
with the ionizing star in its focus and
a lower surface brightness tail (see the 2 cm radio
continuum image of \citeauthor{wood91} \citeyear{wood91}, as well as  
the K-band image of \citeauthor{lumsden96} \citeyear{lumsden96}).

\item The molecular gas traced by H$_2$ shows a similar bright arc
(see Fig.~\ref{fig:vlt:G29}),
but with the peak
displaced `deeper' into the molecular cloud (by 0\farcs93) than the
ionized gas. 

\item The ionized gas shows a steep gradient 
(see Fig.~\ref{fig:vlt:vel}) which goes from $-31$ \kms\ (just in
front of the bright arc) to $+7$ \kms\ (just behind the position of the
star) relative to the molecular cloud velocity (which is about 98 \kms). 
The velocity then drops again to $-4$ \kms\ in
the tail.
 
\item In contrast with the large variations in velocity
observed for Br$\gamma$
across the symmetry axis of G29.96, the H$_2$ emission shows no
significant velocity gradients. Indeed, it seems to be at the velocity
of the molecular cloud.

\item The FWHM of the Br$\gamma$ and H$_2$ lines are significantly
higher in front of the respective bright arcs than in the
tail. Specifically, the linewidths observed for Br$\gamma$
range from $42-46$ \kms\ behind the star to values as high as 62 \kms\
just in front of the arc. In the case of the H$_2$ line, the
linewidth increases from 33 \kms\ in the tail to $\sim 50$ \kms\ in front
of the H$_2$ arc.

\end{enumerate}

Two different models aimed at explaining the formation and evolution
of cometary \HII\ regions and applied to G29.9 appear in the
literature: the bow shock model and the champagne
flow model. Neither of these two models can account for all the 
characteristics observed for G29.96. We critically 
discuss both models.

\subsection{The stellar bow shock model}
\label{sect:vlt:bow}

The morphology of cometary \HII\ regions, in particular their
near-parabolic shape, has been interpreted in terms of a bow
shock \citep{wood89}. In such a model, an OB star moves with
supersonic velocity (of a few \kms) inside a
molecular cloud. The 
interaction of the stellar wind with the ambient medium results in the
formation of a bow shock with a cometary
morphology. Detailed analytical and numerical models have been 
developed and applied to the UCH{\sc ii} region G29.96
\citep{vanburen90,maclow91,vanburen92}.

The structure proposed by \cite{vanburen90} of an UCH{\sc ii}
region trapped in a bow shock is the following. 
The fast stellar wind (with
a terminal velocity in excess of 2000 \kms) rams into
the molecular cloud material, which is moving at supersonic velocity
with respect to the star, resulting in the formation of
a shock wave. The shock occurs
at a radius where the momentum flux in the wind equals the ram
pressure of the ambient medium.
The immediate post-shock gas temperature is $\sim 10^8$ K. At this
temperature, radiative cooling is not very efficient, but cooling
via thermal conduction can lower the temperature very rapidly and consequently,
this layer of shocked wind gas is rather thin \citep{comeron98}. 
As the star moves through the ambient molecular medium, the hot bubble
of shocked wind gas
sweeps up material into a thin parabolic shell. 
Because of the rapid cooling, the wind only transfers its momentum to
the swept-up shell in what is called a ``momentum-driven'' bow-shock. 
This shell traps all or part of the
ionizing flux of the star, forming a thin layer of dense ionized gas
surrounded by shocked molecular gas. The layer of ionized gas has the 
properties of
a typical radiatively excited \HII\ region at a temperature of  
$\sim 10^4$ K. 

In terms of the dynamics, the bow shock models essentially
predict a redshift or blueshift of the ionized
gas relative to the ambient cloud material depending on whether the
star is moving away
from the observer and into the molecular cloud, or towards the
observer. The material in the bow shock is transported (due to a
pressure gradient) from the head to the tail with a velocity that will
not exceed the space velocity of the star by much. Note, however, that
the material inside the bow shock flows in a direction away from the
head to the tail, i.e. at some point it moves in a direction opposite
from the stellar motion.
The maximum velocity difference is then of the order of
$v_{\star}\,{{\rm cos}(i)}$, where 
$v_{\star}$ is the relative velocity of the star through the ISM and
$i$ is the inclination angle. 
As emphasized by \cite{lumsden96}, the velocity structure observed for 
the Br$\gamma$ line can only be explained by the bow shock model
if the molecular cloud
velocity is 67 \kms, rather than the 98 \kms\ obtained  from 
rotational lines 
\citep[e.g.][]{churchwell90b,churchwell92,cesaroni92,bronfman96,olmi99}.
Moreover, if we attribute the difference of $\simeq 32$ \kms\ observed 
in between positions $-7$\arcsec and 0\arcsec\
(see Fig.~\ref{fig:vlt:vel}\,a) to the star-cloud
interaction, then, with the inclination angle estimated from the
morphology of $-45$ degrees \citep{vanburen92}, this translates into a
stellar velocity of $\simeq 45$ \kms.  This is a very high velocity
for a star in a cluster. Now, the difficulty with the stellar velocity 
might be mitigated if the velocity shoulder from 67 to 87 \kms\
observed in Fig.~\ref{fig:vlt:vel}\,a between positions $-7$\arcsec\ and
$-3$\arcsec\ traces motions in the hot core located just 2\arcsec\
west of the bright arc. This hot core is known to have an embedded
protostellar object with a prominent outflow (A. Gibb, private
communication). In that case, the required molecular cloud velocity is 
87 \kms\ and the stellar velocity needed to
explain the velocity difference between positions 
$-3$\arcsec and 0\arcsec\ is only 17 \kms. However, still this
velocity of 87 \kms\ is much less than the observed molecular cloud
velocity of 98 \kms.

The bow shock model provides a natural explanation for the general
morphology of cometary \HII\ regions if the inclination angle is
larger than $\pm 30$ degrees \citep{maclow91,vanburen92}.
As an aside, we can test whether the morphology observed for G29.96  is 
consistent with that of an O5--O6 main sequence star 
(see Sect.~\ref{sect:vlt:sptype}) developing a bow shock.
Van Buren et al. (1990) propose a simple analytical model for
bow shocks in molecular clouds. The standoff distance $l_{\rm sd}$, i.e. the
distance from the star where the momentum flux in the wind equals the
ram pressure of the ambient medium, is given by:

\begin{eqnarray}
l_{\rm sd} ({\rm pc}) = 0.015
\left ( {\dot{m}_{\star} \over {10^{-6} \, {\rm M}_{\sun}/{\rm yr}}}
\right ) ^{1/2} 
\left ( {v_{\rm w} \over {10^{8} \, {\rm cm}/{\rm s}}} \right ) ^{1/2}
\nonumber \\
\left ( {n_{\rm H} \over {10^{5} \, {\rm cm}^{-3}}} \right ) ^{-1/2} 
\left ( {v_{\star} \over {10^{6} \, {\rm cm}/{\rm s}}} \right)^{-1}
\nonumber
\label{eq:vlt:lsd}
\end{eqnarray}

\noindent
where $\dot{m}_{\star}$ is the stellar wind mass-loss rate, $v_{\rm w}$ 
is the terminal velocity of the wind, $n_{\rm H}$ is the number
density of hydrogen nuclei in all forms in the ambient gas and
$v_{\star}$ is the relative velocity of the star through the ISM. 
We have adopted a mean mass per particle, $\mu_{\rm H}$, of 1.4 for
the molecular gas.
We take an O6 star with a metallicity twice solar
(see Sect.~\ref{sect:vlt:he+}) and stellar parameters adopted
from \cite{vacca96}. Following \cite{vink01}, the corresponding
stellar wind
parameters are $\dot{m}_\star = 2.81 \times 10^{-6}$~M$_{\sun}$~yr$^{-1}$ and
$v_{\rm w}= 2951$~\kms. A star of these characteristics moving
supersonically trough the molecular cloud at a velocity of 17 \kms\
will develop a bow shock with a standoff distance of 0.025 pc for a
typical ambient gas density of $10^5$ \cm3. At a distance of 6 kpc and 
for an inclination angle of $-45$ degrees, 
this distance corresponds to a projected separation of 0\farcs6. 
From the observations
(see Sect.~\ref{sect:vlt:results}), the projected standoff
distance would be the offset between the star and the peak of the 
Br$\gamma$ arc, i.e. 2\farcs27. The observations
imply either a much lower stellar velocity (of the order of 5
\kms) or a molecular gas density at the stagnation point
below $10^4$ \cm3.

\subsection{The champagne flow model}
\label{sect:vlt:champagne}

G29.96 has also been interpreted in terms of a champagne flow. 
Champagne flow (blister) models assume that the medium in which a 
massive star is
born is not uniform but has strong density
gradients, which can give rise to an \HII\ region that expands
supersonically  away from the high-density region in a so-called
champagne flow \citep[e.g.][]{tenorio79,bodenheimer79}. 
Theoretical simulations of the expansion of an
\HII\ region in the champagne phase, i.e. when the ionization front
of the nebula crosses a region of strong density gradients, such as the
edge of the molecular cloud, and expands into the lower density
intercloud medium, have been presented by 
\cite{yorke83}. During this champagne flow phase, the \HII\ region
is ionization bounded on
the high-density side (i.e. towards the molecular cloud) and 
density bounded on the side of the outward
champagne flow. 

The morphology of the
ionized gas as predicted by these models
is characterized by a bright compact component and an
extended low brightness component roughly shaped like
an opened fan. This compact component is inconsistent with the
morphology observed for G29.96. At the same time, the opening of the
morphology predicted for the tail region is not evident in G29.96.

In terms of the velocity structure, the numerical calculations
\citep{yorke83} show that the ionized gas 
is accelerated in the direction 
away from the molecular cloud to a relatively high velocity, which can 
attain values in excess of 30 \kms. On the contrary, very little variations
should be observed near the core of the \HII\ region, where the
ionization front should be essentially at the molecular cloud
velocity (98 \kms\ in the case of G29.96). However, we see it at 67
\kms\ (or 87 \kms\ if we attribute the velocity shoulder between
position $-7$\arcsec\ and $-3$\arcsec\ to velocity flows within the
hot core, see Sect.~\ref{sect:vlt:bow}).

\subsection{Final remarks}

Neither of the two competing models can explain the observed velocity
structure for G29.96.  
Partly, this may reflect the complex structure
of this region, where the presence of a hot core with an embedded
young stellar object and an associated outflow may obfuscate the
dynamics. 

Further progress will require the direct determination of the 
stellar velocity from photospheric absorption lines, for instance,
from near-IR spectroscopy at high spectral resolution. We tried to obtain
information on the stellar velocity from the \ion{C}{iv} and
\ion{N}{iii} emission lines observed in our K-band spectrum, but it was
not possible due to the complexity
of these lines together with the limited spectral resolution
($\Delta{\rm v}\sim 33$ \kms). Also, 
a complete mapping of the dynamics of the ionized and molecular gas at both
high spectral and spatial resolution will be essential.

A comparison of the observations with alternative models will also be
important. Several authors, for instance,  have
investigated the interaction of a stellar wind and a clumpy
molecular cloud material \citep[e.g.][]{dyson95,redman96,williams96}. 
In this  mass-loaded model, 
material from dense neutral clumps of gas is added to the ionized
region around a newly formed massive stars; this neutral material
soaks up ionizing photons, slowing down the expansion of the
UCH{\sc ii} region. This mass-load model can explain cometary
regions if the star is located in a density gradient of mass-loading
clumps \citep{dyson98}. G29.96 provides an excellent laboratory to
test this model.

\section{Summary}
\label{sect:vlt:summary}

We presented a high quality, medium resolution K-band spectrum of the
ultracompact region G29.96$-$0.02 obtained with the VLT. The slit was
positioned along the symmetry axis of the nebula. Besides the spectrum
of the embedded ionizing star, the observations included the
emission-line spectrum produced by the ionized gas with sub-arcsec
spatial resolution. The nebular spectrum includes Br$\gamma$, several
helium emission lines, and a molecular hydrogen line.

The study of the relative strength of the Br$\gamma$ and \HeI\ lines
as a function of position along the slit allowed us different lines of
investigations. First, the observed helium lines in 
the nebula have been confronted with the \HeI\ recombination
model. We have shown that the deviations of the transition of
4$^3$S--3$^3$P at 2.1128 \mum\ from the Case B approximation are well
described by the current understanding of the collisional and
radiative transfer effects on the \HeI\ recombination spectrum of
nebulae. The Case B  approximation reproduces the ratios of \HeI\ lines
originating from higher quantum states reasonably well. 
These \HeI\ line ratios indicate a constant electron temperature, 
between 6400 and 7500 K.
Second, the relative strength of the Br$\gamma$ and the singlet transition
of 4$^1$S--3$^1$P at 2.1137~\mum\ indicates that He$^+$/H$^+$ is
practically constant along the slit, with an average abundance of
$0.087 \pm 0.001$; we argue that He is singly ionized throughout the
nebula. This conditions sets a lower limit to the effective
temperature of the star.

The K-band spectrum of the ionizing star, together with the constraint
imposed by the coincidence of the He$^+$ and H$^+$ zones, limit its
spectral type to O5--O6. 
The various observational constraints on the effective temperature of the
ionizing star are reviewed, and they are shown to be in rather good
agreement with
the above spectral type. 

Finally, the variations in peak velocity and width of the Br$\gamma$
and H$_2$ lines allowed us to study the kinematical structure of the 
ionized and molecular gas at higher spatial resolution than in
previous studies. We find that neither the wind bow shock model nor
the champagne flow model are supported by the observations. A firm
constraint on the velocity of the star and a complete mapping of the 
dynamics of the ionized and molecular gas at both
high spectral and spatial resolution will be essential to understand
the formation and evolution of G29.96.

\begin{acknowledgements}
We would like to thank the referee, Dr. P. Conti.
MICE is supported at MPE by DLR (DARA) under grants 50 QI 86108 
and 50 QI 94023.
\end{acknowledgements}


\end{document}